\documentclass[sigconf]{acmart}
\AtBeginDocument{%
  }

\setcopyright{acmlicensed}
\copyrightyear{2026}
\acmYear{2026}
\setcopyright{cc}
\setcctype{by-nc-nd}
\acmConference[CHI EA '26]{Extended Abstracts of the 2026 CHI Conference on Human Factors in Computing Systems}{April 13--17, 2026}{Barcelona, Spain}
\acmBooktitle{Extended Abstracts of the 2026 CHI Conference on Human Factors in Computing Systems (CHI EA '26), April 13--17, 2026, Barcelona, Spain}
\acmDOI{10.1145/3772363.3798377}
\acmISBN{979-8-4007-2281-3/2026/04}

\usepackage{graphicx}
\usepackage{multirow}
\usepackage{caption,subcaption,graphicx}
\usepackage{xcolor}
\usepackage{booktabs}
\usepackage{hyperref}
\begin{document}

%\title{Interactive Rule-Guided AI Failure Detection and Model Adaptation for Human-AI Collaborative Decision-Making Systems}
\title{From Accuracy to Readiness: Metrics and Benchmarks for Human–AI Decision-Making}

\author{Min Hun Lee}
\email{mhlee@smu.edu.sg}
\orcid{0000-0003-3506-8972}
\affiliation{%
  \institution{Singapore Management University}
  \country{Singapore}
}
\renewcommand{\shortauthors}{Lee.}

\begin{abstract}
Artificial intelligence (AI) systems are deployed as collaborators in human decision-making. Yet, evaluation practices focus primarily on model accuracy rather than whether human-AI teams are prepared to collaborate safely and effectively. Empirical evidence shows that many failures arise from miscalibrated reliance, including overuse when AI is wrong and underuse when it is helpful. 

This paper proposes a measurement framework for evaluating human-AI decision-making centered on team readiness. We introduce a four-part taxonomy of evaluation metrics spanning outcomes, reliance behavior, safety signals, and learning over time, and connect these metrics to the Understand–Control–Improve (U–C–I) lifecycle of human-AI onboarding and collaboration. 

By operationalizing evaluation through interaction traces rather than model properties or self-reported trust, our framework enables deployment-relevant assessment of calibration, error recovery, and governance. We aim to support more comparable benchmarks and cumulative research on human–AI readiness, advancing safer and more accountable human–AI collaboration.
\end{abstract}






\maketitle

\section{Introduction}
Artificial intelligence (AI) systems are increasingly deployed as collaborators rather than autonomous decision-makers, supporting human judgment in high-stakes domains such as healthcare \cite{cai2019human,lee2020co,lee2021human,wang2021brilliant} and public services \cite{kuo2023understanding,zavrvsnik2020criminal}. In these settings, AI systems increasingly shape how people interpret evidence, calibrate confidence, allocate responsibility, and ultimately make decisions \cite{holstein2023toward,lee2021human,lai2021towards,cai2021onboarding}. 

Over the past several years, empirical Human--AI Interaction (HAI) research has demonstrated that model performance alone is insufficient for safe and effective human--AI collaboration: even highly accurate systems can yield worse human–AI outcomes when users follow incorrect advice, ignore correct advice, or apply inconsistent intervention strategies under uncertainty \citep{he2023stated,bansal2021most,buccinca2021trust,chen2023understanding,lai2021towards}. Complementing these findings, research on accountable and trustworthy AI emphasizes governance mechanisms, such as oversight, contestability, auditing, and responsibility across deployment \citep{raji2020closing,novelli2024accountability,mokander2022algorithmic,kaur2022trustworthy,vashney2022trustworthy,toreini2020relationship,lee2026ruleedit}. Meanwhile, explainable AI (XAI) and interactive ML research has proposed many mechanisms—feature attributions, examples, counterfactuals, rules, and uncertainty estimates—to make model behavior intelligible \citep{ribeiro2016should,arya2019one,doshi2017towards,wang2019designing,rudin2019stop,crisan2022interactive,guo2022building,kulesza2011oriented}. However, empirical evidence across HAI and XAI suggests these techniques do not reliably improve decision quality by default. Instead, their effects depend on task context, user expertise, timing, and interactions with human intuition and confidence \citep{chen2023understanding,he2023stated,lai2021towards,buccinca2021trust,lee2023understanding}.

Despite progress on mechanisms, evaluation practices remain misaligned with how human–AI systems fail in practice during real-world deployment. Many studies emphasize model accuracy, explanation fidelity, or self-reported trust \cite{lai2021towards,guo2024decision,rudin2019stop,ghassemi2021false}, implicitly assuming these proxies reflect whether users are ready to collaborate with AI safely and effectively. Yet, trust often poorly predicts reliance behavior, and explanations can increase overreliance by providing a false sense of certainty or legitimacy \citep{lai2021towards,lee2023understanding,chen2023understanding,bansal2021most,ghassemi2021false}. Consequently, real-world failures persist not only due to model error, but due to miscalibrated human reliance—overreliance when AI is wrong, underuse when AI is helpful, and brittle “local” adaptations that do not generalize across cases \citep{he2023stated,buccinca2021trust,lee2023understanding,chen2023understanding,ghassemi2021false}. Critically, these failure modes are often invisible when evaluation reports only accuracy, perceived trust, or explanation satisfaction.

\begin{figure*}[htp]
  \centering
\includegraphics[width=0.9\linewidth]%
  {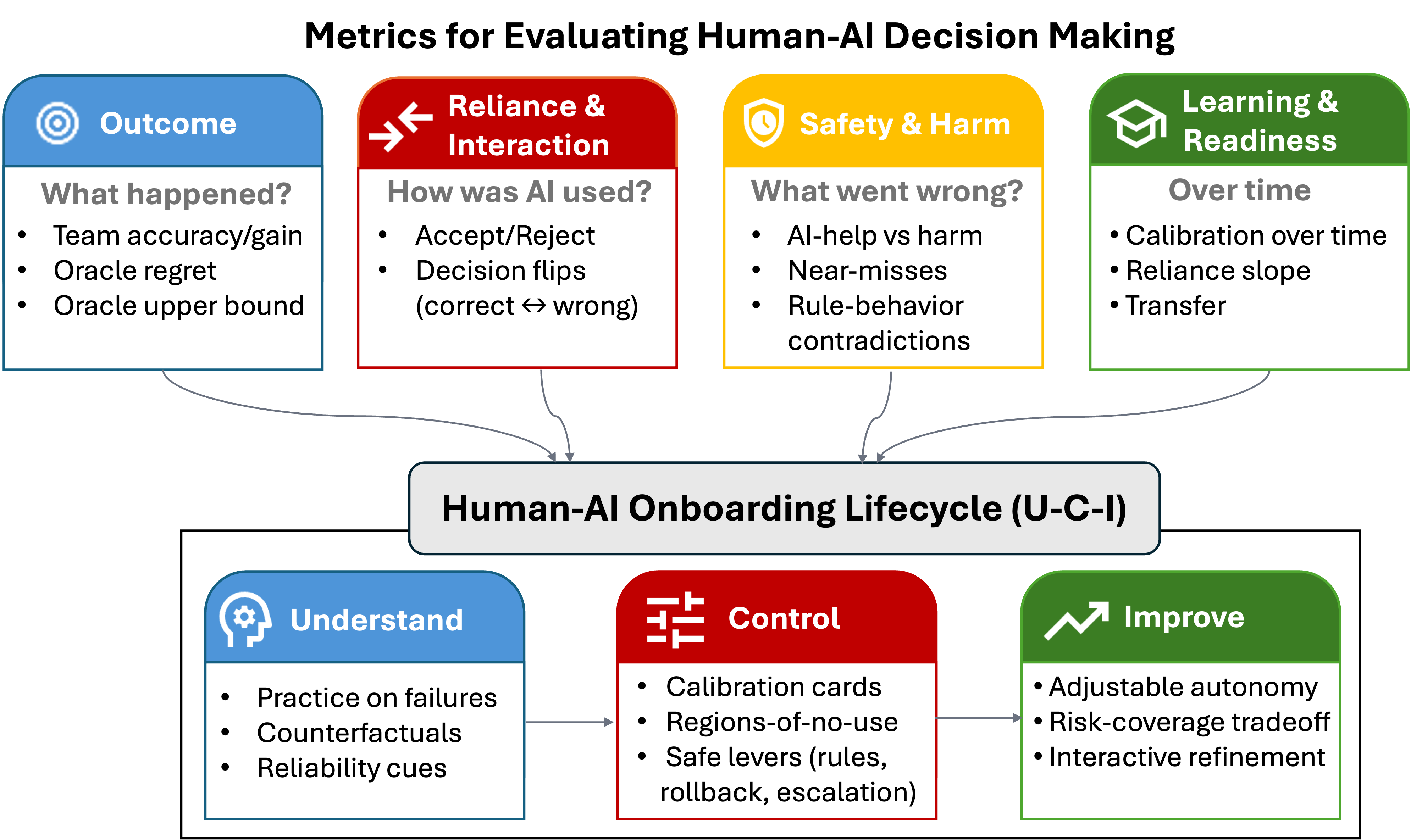}
\caption{We propose a four-part taxonomy of metrics for human–AI onboarding (top) and show how each metric family becomes observable and actionable across the human-AI onboarding lifecycle (Understand–Control–Improve) (bottom).}
\Description{A two-level framework linking four metric families for human–AI decision-making (Outcome, Reliance & Interaction, Safety & Harm, and Learning & Readiness) to three stages of the human–AI onboarding lifecycle: Understand, Control, and Improve.}
\label{fig:interface}
\end{figure*}
%Figure 1: A two-level framework linking four metric families for human–AI decision-making (Outcome, Reliance & Interaction, Safety & Harm, and Learning & Readiness) to three stages of the human–AI onboarding lifecycle: Understand, Control, and Improve.

In this paper, we argue that resolving this gap requires shifting evaluation from “how good is the model?” to “how ready is the human–AI team?”: whether users can recognize failures, calibrate reliance, and remain accountable under realistic constraints \citep{lai2021towards,bansal2021most,buccinca2021trust,chen2023understanding}. We focus on onboarding, calibration, and governance as the early-deployment phase where reliance patterns are formed and where many downstream failures originate \citep{cai2019hello,mitchell2019model}. Building on this direction, our work reframes onboarding as a measurable learning intervention organized around \textbf{Understand–Control–Improve (U–C–I)},  extending recent work on AI onboarding and explanation-supported learning for clinical decision-making \citep{lee2024interactive,lee2024improving}. We treat onboarding broadly as the process through which users learn to work effectively with AI systems in real decision-making settings. In \textbf{Understand}, users develop mental models of model behavior, boundary conditions, and failure modes through structured practice on curated failure sets and counterfactual examples that reveal how small input changes can flip predictions \citep{bansal2019beyond,verma2020counterfactual,lee2023understanding}. In \textbf{Control}, users learn how to calibrate reliance and apply safe interventions using lightweight supports, such as \textit{calibration cards}, artifacts that summarize when AI predictions are reliable or unreliable (e.g. “when to trust”, “when to double-check”) \cite{lee2026ruleedit}, common failure modes \cite{cai2019hello,lee2024improving}, and recommended operating points (e.g. thresholds or escalation rules) \cite{lee2025towards}, alongside {regions-of-no-use}, contexts where AI recommendations should not be trusted \cite{lee2024interactive}, and safe levers, user-facing controls (e.g. rule/threshold edits) that allow bounded intervention in AI behavior with preview, rollback, and audit trails to support contestability \cite{kulesza2015principles,amershi2019guidelines,lyons2021conceptualising,lee2021human,alfrink2023contestable,lee2026ruleedit} and accountability \citep{mitchell2019model,raji2020closing,crisan2022interactive}. In \textbf{Improve}, teams iteratively refine collaboration strategies and governance policies using feedback from newly observed failures to update training content, thresholds, and governance practices \citep{bansal2021most,lai2021towards,lee2021human,kulesza2015principles,lee2026ruleedit}.

We organize these measures around the Understand–Control\\–Improve (U–C–I) lifecycle. U–C–I describes when key capabilities in human–AI collaboration develop: users first learn model behavior and limitations (Understand), then calibrate how and when AI should be used in practice (Control), and finally refine collaboration strategies and governance policies over time (Improve). The four metric families describe what should be measured across this lifecycle: Outcome metrics evaluate decision quality, Reliance and Interaction metrics capture how AI advice is adopted or rejected, Safety and Harm metrics identify high-risk collaboration failures, and Learning and Readiness metrics measure how these behaviors evolve over repeated use. Together, the taxonomy makes the U–C–I lifecycle observable, enabling evaluation of how human–AI collaboration evolves over time.

Prior work has proposed a variety of measures for evaluating human–AI interaction, including trust, reliance, agreement with model predictions, and decision accuracy \cite{green2019disparate,lai2021towards,schemmer2023appropriate,bansal2021does}. However, these measures are often studied in isolation and therefore do not capture the full lifecycle of human–AI collaboration. We synthesize these existing constructs into four complementary metric families: outcome quality, reliance behavior, safety and harm signals, and learning over time. These categories reflect four practical questions that arise when deploying AI decision-support systems: What happened? How was the AI used? What went wrong? And how does collaboration evolve over time?

%Building on this lifecycle framing, we propose a taxonomy of metrics for human–AI onboarding and decision-making that makes onboarding measurable and evaluation results comparable across studies and domains. 
Building on this framing, we further specify how these metrics can be computed directly from observable interaction traces rather than inferred attitudes or model properties. Our taxonomy spans four complementary classes.

\begin{enumerate}
    \item Outcome metrics: capture decision quality beyond raw model accuracy, such as team gain and avoidable error (e.g. regret relative to the best achievable human–AI decision), reflecting whether AI involvement ultimately improves or degrades outcomes \citep{bansal2021most,guo2024decision}.
    \item Reliance \& Interaction metrics: characterize how AI advice shapes human judgments, including accept-on-wrong, changed-to-wrong, override frequency and timing, and reliance slope, which operationalize behavioral calibration and sensitivity to AI correctness \citep{buccinca2021trust,chen2023understanding,lai2021towards,lee2023understanding}.
    \item Safety \& Harm metrics: attribute risk to AI influence and governance breakdowns rather than human error alone, including AI-induced harm, near-misses, contradictions between rules and behavior, and rollback or escalation events \citep{raji2020closing,ghassemi2021false}.
    \item Learning \& Readiness metrics: assess whether onboarding produces durable skill, such as failure identification, explanation comprehension, and retention or transfer across cases, tasks, or model versions \citep{cai2019hello,holstein2023toward}.
\end{enumerate}

These four metric families can be instantiated across a wide range of decision-support settings. For example, in a clinical triage system \cite{lee2025towards}, outcome metrics measure the accuracy of the final human–AI decision, reliance metrics capture how often clinicians accept or override AI recommendations, safety metrics detect harmful deferrals to incorrect AI predictions, and learning metrics track how reliance evolves across repeated cases.

These metrics are not standalone statistics. They are computed from decision traces (e.g. accept, override, change), error attribution (AI-influenced versus independent errors), and learning signals (e.g. pre/post onboarding probes, time-to-calibration, cross-case transfer). As a result, each metric class maps naturally to stages of the \textbf{Understand–Control–Improve (U–C–I)} onboarding lifecycle, where it becomes both observable (during interaction) and actionable (through training, control levers, or governance interventions). This structure moves evaluation beyond accuracy and trust toward cumulative, deployment-relevant evidence of human–AI readiness.

This framework surfaces a measurement and benchmarking agenda for the human--AI interaction community: 
\begin{itemize}
    \item {When does a user become “AI-ready”?}
    \item Which reliance and harm metrics generalize across domains?
    \item How should governance be evaluated in use—beyond documentation—through behaviors such as contestation, rollback, escalation, and auditability?
\end{itemize}
 
Answering these questions would enable cumulative science and more deployment-relevant evidence for safe human–AI collaboration. We argue that progress in human–AI collaboration requires shifting from evaluating models in isolation to evaluating human–AI teams, and from reporting isolated metrics to developing benchmarkable measures of readiness, calibration, and governance.

\textbf{Positioning:} Our work complements prior frameworks for measuring reliance in human--AI systems \cite{guo2024decision} and surveys of human--AI decision-making metrics \cite{lai2021towards}. While these works catalog existing measures or analyze reliance behavior, we focus on evaluation during onboarding and early deployment, where reliance patterns are formed and many downstream failures originate. We therefore propose a structured taxonomy of evaluation metrics and map these metrics to actionable stages in the Understand–Control–Improve (U–C–I) lifecycle.

\textbf{Contribution:} We contribute a \emph{unified, traced-based evaluation framework} for human--AI readiness: 
\begin{itemize}
    \item A metric taxonomy spanning outcomes, reliance, harm, learning
    \item Trace-based metric definitions grounded in interaction logs
    \item A mapping from metrics to actionable \textbf{U--C--I} design interventions (Tables~\ref{tab:metric_mapping_uci_part1}--\ref{tab:metric_mapping_uci_part2}; Appendix~\ref{appendix:metrics}).
\end{itemize}

%to Guo et al.’s decision-theoretic framework for measuring reliance , which is valuable for auditing outcomes but does not address how practitioners are onboarded into safe teamwork or which interventions help users learn failure modes and calibrate reliance over time. It also complements Lai et al.’s survey \cite{lai2021towards}, which synthesizes empirical studies and reported metrics; in contrast, we focus on the onboarding phase and contribute a prescriptive onboarding + governance approach (U–C–I) that links workflow mechanisms to evaluation targets.

%We contribute an \emph{operational evaluation scaffold} for human--AI onboarding and decision-making. Synthesizing evidence across  XAI and responsible AI, we (1) propose a metric taxonomy that disambiguates outcomes, reliance, harm, and learning; (2) define trace-based measures grounded in observable interaction behavior (not self-reported attitudes or model properties); and (3) map metrics to actionable design interventions across the \textbf{U--C--I} onboarding lifecycle (Tables~\ref{tab:metric_mapping_uci_part1}--\ref{tab:metric_mapping_uci_part2}; Appendix~\ref{appendix:metrics}). Our goal is to enable cumulative, comparable, and deployment-relevant evaluation of human--AI onboarding.

\section{Why Accuracy Alone Is Insufficient}

\subsection{Why Current Evaluation Fails}
Despite rapid advances in model performance, many failures of human–AI systems arise after deployment, during everyday use in real workflows. A growing body of HAI research suggests that this gap is not primarily due to insufficient model accuracy, but to a mismatch between how systems are evaluated and how they are actually used \citep{lai2021towards,guo2024decision}. In practice, AI systems are embedded in time pressure, institutional norms, accountability structures, and evolving user strategies, which are rarely reflected in standard evaluation protocols. The following three evaluation assumptions illustrate this mismatch.

\subsubsection{Accuracy $\neq$ Safety}
Accuracy measures whether a model’s prediction matches ground truth, but it does not capture the quality of human–AI decisions. In high-stakes settings, such as healthcare, multiple studies show that users may change initially correct judgments to incorrect ones after seeing AI advice, a phenomenon often referred to as AI-induced error or automation bias \citep{bansal2021most,buccinca2021trust,lee2023understanding}. These errors are invisible in standard accuracy (e.g. AUROC, or F1 metrics), which treat AI outputs as independent of human behavior. Moreover, accuracy does not distinguish between errors that users recognize and recover from versus errors that propagate silently into downstream decisions, documentation, or treatment plans \citep{ghassemi2021false,buccinca2021trust}. As a result, systems that appear high-performing in offline benchmarks may still increase harm when integrated into real workflows where AI advice shapes human judgment.

\subsubsection{Trust $\neq$ Reliance}
Trust is frequently measured through post-task surveys or Likert-scale questionnaires, yet behavioral evidence consistently shows weak alignment between reported trust and actual reliance \citep{lai2021towards,buccinca2021trust,chen2023understanding,lee2023understanding}. Users may report low trust while still following AI recommendations under time pressure, cognitive load, or organizational expectations. Conversely, users may report high trust while selectively ignoring AI advice in critical or ambiguous cases \citep{bansal2021most,lee2023understanding}. This disconnect arises because trust captures attitudes, whereas reliance reflects situated behavior under constraints—including workload, accountability, and perceived risk. Evaluations that rely primarily on trust scores therefore miss when, how, and why users defer to or override AI advice in practice, obscuring important safety and governance concerns.

\subsubsection{Performance $\neq$ Readiness}
High task performance during evaluation does not imply that users are prepared for real-world deployment. Short-term performance gains can mask brittle strategies, such as copying AI outputs without understanding underlying uncertainty or failure modes \citep{he2023stated,buccinca2021trust,lee2023understanding}. In contrast, readiness depends on whether users can recognize when AI is likely wrong, interpret confidence and uncertainty appropriately, and recover from errors when they occur \citep{cai2019hello,holstein2023toward,lee2023understanding,salimzadeh2024dealing,prabhudesai2023understanding}. These capacities (e.g. failure detection, uncertainty interpretation, and error recovery) are rarely measured explicitly, yet they determine whether human–AI systems remain safe over time, under distribution shift, and as models or workflows evolve \citep{lai2021towards,ghassemi2021false}.

Together, these gaps point to a fundamental mismatch: we often evaluate AI systems as artifacts optimized for predictive performance, but deploy them as teammates embedded in human workflows. Addressing this mismatch requires evaluation frameworks that capture not only what the AI predicts, but how humans learn to work with it \cite{cai2019hello}, rely on it \cite{lai2021towards,guo2024decision}, and govern it over time.

\subsection{Reframing Onboarding as a Measurable Process}
To address this mismatch, we reframe onboarding not as documentation, demos, or one-off training, but as a \emph{measurable learning intervention} that prepares users to collaborate with AI safely in real workflows \citep{cai2019hello,holstein2023toward}. Drawing on prior work in human--AI collaboration, explainable AI, learning-by-doing, and AI onboarding for clinical decision-making \citep{lai2021towards,buccinca2021trust,chen2023understanding,cai2019hello,lee2024interactive,lee2024improving}, we conceptualize onboarding as the process through which users acquire durable skills for forming accurate mental models of AI reliability, calibrating reliance, and enacting accountability under realistic constraints.

Effective onboarding supports at least four competencies. First, users learn to \textbf{detect reliability boundaries}: when AI is likely correct or incorrect rather than assuming uniform performance across cases, contexts, or subpopulations \citep{cai2019hello}. Second, users learn to \textbf{calibrate reliance}, adjusting when to accept, question, or override AI advice based on evidence and uncertainty cues \citep{bansal2021most,buccinca2021trust,chen2023understanding}. Third, users learn to \textbf{exercise safe control and contestability}, including how to intervene \cite{kulesza2015principles,guo2022building}, escalate ambiguous cases, and use rollback or audit mechanisms when AI advice conflicts with domain judgment or policy requirements \citep{raji2020closing,mitchell2019model,mokander2022algorithmic}. Fourth, users learn to \textbf{understand delegation and autonomy}, recognizing how responsibility shifts between human and AI under different operating modes (e.g., decision support vs.\ selective deferral) and how these choices affect outcomes and accountability \citep{bansal2021most,guo2024decision,wilder2020learning,holstein2023toward,lee2025towards}.

These abilities cannot be inferred from model properties or self-reported attitudes alone; they must be measured behaviorally through \emph{interaction traces over time} (e.g. acceptance/override patterns, sensitivity to AI correctness, failure detection rates, and recovery actions across cases and changing conditions) \citep{lai2021towards,chen2023understanding,holstein2023toward}.

\subsection{A Taxonomy of Metrics for Human--AI Onboarding \& Decision-Making}
Building on empirical findings across healthcare AI onboarding, decision-support evaluation, uncertainty-aware delegation, and accountable AI systems, we propose a taxonomy of metrics that capture complementary aspects of onboarding and collaboration \citep{cai2019hello,lai2021towards,guo2024decision,raji2020closing,lee2023understanding,prabhudesai2023understanding}. Our taxonomy separates four evaluation questions: \emph{what happened}, \emph{how AI was used}, \emph{what went wrong}, and \emph{what changed over time}---dimensions often conflated or omitted in prior evaluations \citep{lai2021towards,buccinca2021trust,ghassemi2021false}. Full metric definitions and equations are provided in Appendix~\ref{appendix:metrics}.

\subsubsection{Outcome Metrics (What happened?)}
Outcome metrics capture the quality of final human--AI decisions beyond raw model correctness, reflecting whether AI involvement ultimately improves or degrades task outcomes \citep{bansal2021most,guo2024decision}. We report: 
(i) \textbf{team gain} relative to human-only and AI-only baselines, and 
(ii) \textbf{regret\_best}, which quantifies avoidable error relative to an oracle that selects the better of the initial human decision and AI prediction per case \citep{guo2024decision}. 
We further distinguish \textbf{error recovery vs.\ error amplification}, separating cases where AI helps users correct initial mistakes from cases where AI induces harm that would not otherwise occur \citep{he2023stated,ghassemi2021false}. 
\textbf{Oracle best accuracy} is treated as a reference upper bound rather than an operational target, enabling diagnosis of collaboration failures distinct from model limitations  \citep{guo2024decision} (Appendix~\ref{appendix:metrics}).

\subsubsection{Reliance \& Interaction Metrics (How was AI used?)}
Reliance metrics characterize \emph{how} AI advice shapes human decisions, operationalizing behavioral calibration rather than subjective attitudes \citep{buccinca2021trust,chen2023understanding,lai2021towards}. We track: 
(i) \textbf{accept-on-wrong} (agreeing with incorrect AI), 
(ii) \textbf{changed-to-wrong} (switching from a correct human judgment to an incorrect final decision after seeing AI), 
(iii) \textbf{override frequency and timing}, and 
(iv) \textbf{local vs.\ global update asymmetry} (i.e. whether users treat a failure as case-specific or revise their broader mental model of AI reliability) \citep{wang2023watch,liao2023ai}. These measures expose overreliance, underuse, and brittle strategies that are invisible in aggregate accuracy \citep{bansal2021most} (Appendix~\ref{appendix:metrics}).

\subsubsection{Safety \& Harm Metrics (What went wrong?)}
Safety metrics attribute harm to AI influence and governance breakdowns rather than human error alone \citep{raji2020closing,mokander2022algorithmic,ghassemi2021false}. We include:
(i) \textbf{AI-harm} (cases where AI causes a correct initial human decision to become wrong), 
(ii) \textbf{near-misses} (high-risk disagreements narrowly avoided), and 
(iii) \textbf{governance-in-use signals} such as contradictions between rules and behavior, rollback events, and escalation actions. 
These metrics operationalize accountability as enacted behavior rather than documentation alone \citep{raji2020closing,mitchell2019model} (Appendix~\ref{appendix:metrics}).

\subsubsection{Learning \& Onboarding Metrics (What changed over time?)}
Learning metrics assess whether onboarding produces durable skill \citep{cai2019hello} rather than transient performance gains. We measure: 
(i) \textbf{calibration gap} (confidence vs.\ correctness), 
(ii) \textbf{reliance slope} (acceptance sensitivity to AI correctness), 
(iii) \textbf{stability under distribution shift}, and 
(iv) \textbf{transfer} across tasks, cases, or model versions. These targets operationalize ``AI readiness'' as a behavioral capability that persists beyond a single evaluation outcome \citep{lai2021towards,buccinca2021trust,guo2024decision} (Appendix~\ref{appendix:metrics}).

In operational settings, many of these metrics can be computed directly from interaction logs that record initial human decisions, AI recommendations, and final outcomes. In large-scale deployments, collecting these signals may require event-logging infrastructure similar to observability pipelines used in production ML systems. When ground-truth labels are delayed or expensive, practitioners may estimate some metrics through sampling strategies or proxy signals such as disagreement events or escalation rates. In privacy-sensitive settings, behavioral traces should be collected with appropriate aggregation and anonymization mechanisms.

\subsubsection{Calibration \& Governance as First-Class Targets}
Across domains, outcomes depend less on raw predictive accuracy and more on whether users \emph{calibrate reliance}, accepting AI when it is likely correct and overriding it when it is likely wrong \citep{bansal2021most,lai2021towards,buccinca2021trust}. Even highly accurate systems can degrade team performance when users over-rely on incorrect advice or fail to intervene at critical moments \citep{ghassemi2021false,lai2021towards,buccinca2021trust}. Thus, calibration should be treated as a \emph{primary} evaluation target (e.g. accept-on-wrong, changed-to-wrong, reliance slope, calibration gap), not a byproduct of explainability.

Governance mechanisms (e.g. model cards, audit trails, policies) are necessary but insufficient on their own: accountability is enacted through everyday interaction, including how users contest AI, justify overrides, escalate cases, or rollback edits \citep{mitchell2019model,raji2020closing,mokander2022algorithmic,lee2026ruleedit}. Behavioral signals such as rollback frequency, escalation behavior, contradiction detection, and intervention latency provide empirical evidence of ``governance in use,'' enabling assessment beyond documentation \citep{raji2020closing,ghassemi2021false}.

\subsubsection{Open Benchmarking Questions}
Taken together, our framework raises foundational benchmarking questions for the HCI and HAI community:
\begin{itemize}
    \item \textbf{When is a user “AI-ready”?} What behavioral criteria indicate readiness for deployment, beyond short-term task performance?
    \item \textbf{Which onboarding metrics generalize across domains?} Which measures of reliance, learning, and harm are robust to task context, expertise, and institutional setting?
    \item \textbf{How should governance mechanisms be evaluated empirically?} What behavioral signals best capture contestability, accountability, and safe intervention in use?
    \item \textbf{What should standardized human–AI benchmarks include beyond accuracy?} How can benchmarks reflect calibration, error recovery, and governance rather than prediction alone?
\end{itemize}
Addressing these questions is essential for cumulative, comparable, and deployment-relevant progress in human–AI collaboration research.

\section{Discussion and Conclusion}
This paper positions \emph{measurement} rather than algorithmic novelty as a central bottleneck for safe and accountable AI deployment. By shifting evaluation toward calibration, learning, and governance, the proposed framework aims to support: (i) comparable evaluation across studies and domains, (ii) principled design of onboarding interventions grounded in learning outcomes, and (iii) policy-relevant assessment of AI governance as enacted in practice. In addition, we provide an agenda for future CHI workshops, surveys, benchmarks, and research programs focused on human–AI teaming rather than model-centric performance. 
\textbf{As a limitation, this taxonomy should be understood as a starting point rather than a finalized standard}: it synthesizes recurring measures and highlights gaps, and it will require community iteration, domain-specific validation, and refinement as new evidence and deployment contexts emerge.
If we do not measure onboarding, calibration, and harm, we cannot claim that human–AI systems are ready for real-world collaboration. This work proposes a shared measurement agenda for evaluating human–AI teams—not as tools, but as socio-technical systems whose safety and effectiveness emerge through interaction over time. This framework provides a foundation for future evaluation protocols, benchmark design, and shared measurement standards for human–AI collaboration across domains.

\begin{acks}
This research was supported by the Resilient Workforces Institute, Singapore Management University, under the SMU Seed Fund (Grant ID: 2026-6026IR-25T040-SMUIRNYXXX), and by the Ministry of Education, Singapore under its Academic Research Fund Tier 2 (MOE-T2EP20223-0007). Any opinions, findings and conclusions or recommendations expressed in this material are those of the authors and do not reflect the views of the Ministry of Education, Singapore.
\end{acks}

\bibliographystyle{ACM-Reference-Format}
\bibliography{main}

\newpage
\appendix
\section{Appendix: Detailed Metrics (Organized by the Four-Part Taxonomy)}\label{appendix:metrics}

Let $\{(y_j, h_{0j}, a_j, h_{1j}, c_j, t_j)\}_{j=1}^N$ denote $N$ decision instances, where $y_j$ is the ground truth, $h_{0j}$ the participant’s initial decision, $a_j$ the AI prediction, $h_{1j}$ the participant’s final decision after viewing AI output, $c_j$ the participant’s reported confidence (if available), and $t_j$ ~~timestamps or interaction events recorded in system logs.

We organize metrics by \emph{what happened}, \emph{how AI was used}, \emph{what went wrong}, and \emph{what changed over time}, following prior analyses of human–AI reliance and collaboration behavior \citep{lai2021towards,buccinca2021trust,lee2023understanding,guo2024decision}.

\subsection{Outcome Metrics (What happened?)}

\paragraph{Accuracies and team gains.}
These metrics describe decision quality at the human, AI, and human--team level.

\begin{itemize}
    \item \textbf{Human accuracy} ($\mathrm{Acc}_{h0}$): proportion of cases correctly solved by the human before seeing AI.
    \[
    \mathrm{Acc}_{h0}=\frac{1}{N}\sum_{j=1}^N \mathbb{I}[h_{0j}=y_j]
    \]
    \item \textbf{AI accuracy} ($\mathrm{Acc}_{ai}$): proportion of cases correctly predicted by the AI.
    \[
    \mathrm{Acc}_{ai}=\frac{1}{N}\sum_{j=1}^N \mathbb{I}[a_j=y_j]
    \]
    \item \textbf{Team accuracy} ($\mathrm{Acc}_{team}$): proportion of cases where the final human–AI decision is correct.
    \[
    \mathrm{Acc}_{team}=\frac{1}{N}\sum_{j=1}^N \mathbb{I}[h_{1j}=y_j]
    \]
    \item \textbf{TeamGain vs.\ Human}: improvement (or degradation) from human-only decisions to AI-assisted decisions.
    \[
    \mathrm{Acc}_{team}-\mathrm{Acc}_{h0}
    \]
    \item \textbf{TeamGain vs.\ AI}: improvement (or degradation) from AI-only predictions to the final team decision.
    \[
    \mathrm{Acc}_{team}-\mathrm{Acc}_{ai}
    \]
\end{itemize}

\paragraph{Oracle upper bound and regret.}
These metrics separate model limitations from collaboration failures.
Define the oracle-correct indicator as:
\[
\mathrm{Oracle}_j=\mathbb{I}\big[(h_{0j}=y_j)\lor(a_j=y_j)\big].
\]
\begin{itemize}
    \item \textbf{Oracle best accuracy} ($\mathrm{Acc}_{oracle}$): upper bound on achievable team performance if one always selected the correct agent.
    \[
    \mathrm{Acc}_{oracle}=\frac{1}{N}\sum_{j=1}^N \mathrm{Oracle}_j
    \]
    \item \textbf{Regret\_best}: proportion of avoidable errors where the team fails despite at least one agent being correct \citep{guo2024decision}.
    \[
    \frac{1}{N}\sum_{j=1}^N
    \left(\mathrm{Oracle}_j-\mathbb{I}[h_{1j}=y_j]\right)
    \]
\end{itemize}

\paragraph{Error recovery vs.\ error amplification (derived outcome effects).}
Beyond aggregate accuracy and regret, we distinguish whether AI involvement helps users recover from errors or amplifies them. \emph{Error recovery} refers to cases where an initially incorrect human decision becomes correct after viewing AI output, while \emph{error amplification} refers to cases where a correct initial human decision becomes incorrect due to AI influence. These outcome-level effects are not captured by accuracy or regret alone, but are critical for assessing whether AI improves or degrades real-world decision quality. We operationalize error recovery and amplification through the help--harm decomposition (AI-help vs.\ AI-harm) and complementary decision-change metrics (ChangedToRight vs.\ ChangedToWrong), defined in subsequent sections of the appendix.

% -----------------------
\subsection{Reliance \& Interaction Metrics (How was AI used?)}

\paragraph{Reliance conditioned on AI correctness.}
These metrics capture behavioral reliance patterns, including appropriate reliance as well as over- and under-reliance.\\

Let $\mathcal{C}=\{j:a_j=y_j\}$ and $\mathcal{W}=\{j:a_j\neq y_j\}$.
\begin{itemize}
    \item \textbf{Accept-on-correct}: tendency to follow AI when it is correct.
    \[
    \Pr(h_1=a\mid a=y)=\frac{1}{|\mathcal{C}|}\sum_{j\in\mathcal{C}}\mathbb{I}[h_{1j}=a_j]
    \]
    \item \textbf{Reject-on-wrong}: ability to reject incorrect AI advice.
    \[
    \Pr(h_1\neq a\mid a\neq y)=\frac{1}{|\mathcal{W}|}\sum_{j\in\mathcal{W}}\mathbb{I}[h_{1j}\neq a_j]
    \]
    \item \textbf{Reject-on-correct}: unnecessary rejection of correct AI advice.
    \[
    \Pr(h_1\neq a\mid a=y)
    \]
    \item \textbf{Accept-on-wrong}: overreliance on incorrect AI predictions.
    \[
    \Pr(h_1=a\mid a\neq y)
    \]
\end{itemize}

\paragraph{Decision-change behaviors.}
These metrics distinguish beneficial from harmful decision updates.
\begin{itemize}
    \item \textbf{Changed}: proportion of cases where the participant changes an initial decision after seeing AI.
    \[
    \frac{1}{N}\sum_{j=1}^N\mathbb{I}[h_{1j}\neq h_{0j}]
    \]
    \item \textbf{ChangedToRight}: beneficial changes from incorrect to correct.
    \[
    \frac{1}{N}\sum_{j=1}^N
    \mathbb{I}[h_{1j}\neq h_{0j}]
    \mathbb{I}[h_{0j}\neq y_j]
    \mathbb{I}[h_{1j}=y_j]
    \]
    \item \textbf{ChangedToWrong}: harmful changes induced after seeing AI.
    \[
    \frac{1}{N}\sum_{j=1}^N
    \mathbb{I}[h_{1j}\neq h_{0j}]
    \mathbb{I}[h_{0j}=y_j]
    \mathbb{I}[h_{1j}\neq y_j]
    \]
\end{itemize}

%ChangedToRight and ChangedToWrong provide a complementary behavioral view of recovery vs.\ amplification via decision changes.

\paragraph{Calibration and timing.}
\begin{itemize}
    \item \textbf{Reliance slope}: sensitivity of reliance to AI correctness; higher values indicate better calibration \citep{buccinca2021trust}.
    \[
    \Pr(h_1=a\mid a=y)-\Pr(h_1=a\mid a\neq y)
    \]
    \item \textbf{Intervention latency}: average time taken to confirm or override AI, reflecting hesitation and contesting AI recommendations.
    \[
    \frac{1}{N}\sum_{j=1}^N
    (t^{\text{confirm/override}}_j - t^{\text{AI}}_j)
    \]
\end{itemize}

\paragraph{Local vs.\ global update asymmetry.}
To distinguish case-specific reactions from durable belief updates about AI reliability, we measure whether responses to an AI failure generalize beyond the current instance. \emph{Local updates} are reflected by isolated overrides or rejections confined to the current case, whereas \emph{global updates} manifest as systematic changes in reliance behavior on subsequent cases. We operationalize global updating by comparing reliance metrics (e.g., accept-on-wrong, reliance slope) before versus after observed AI failures, and quantify update asymmetry as the degree to which behavior changes persist across subsequent cases rather than reverting immediately. This framing aligns with evidence that user behavior can shift under model/explanation updates and with calls to study transparency via user mental models over time \citep{wang2023watch,liao2023ai}.

% -----------------------
\subsection{Safety \& Harm Metrics (What went wrong?)}

\paragraph{Help--harm decomposition.}
These metrics attribute outcome changes to AI influence rather than overall accuracy alone.

\begin{itemize}
    \item \textbf{AI-help}: cases where AI corrects an initially wrong human decision.
    \[
    \frac{1}{N}\sum_{j=1}^N
    \mathbb{I}[h_{0j}\neq y_j \land h_{1j}=y_j]
    \]
    \item \textbf{AI-harm}: cases where AI causes a correct human decision to become wrong.
    \[
    \frac{1}{N}\sum_{j=1}^N
    \mathbb{I}[h_{0j}=y_j \land h_{1j}\neq y_j]
    \]
    \item \textbf{Missed-help}: failures to adopt correct AI advice when the human is wrong.
    \[
    \frac{1}{N}\sum_{j=1}^N
    \mathbb{I}[h_{0j}\neq y_j \land a_j=y_j \land h_{1j}\neq a_j]
    \]
    \item \textbf{Correct-ignore}: appropriate rejection of incorrect AI advice.
    \[
    \frac{1}{N}\sum_{j=1}^N
    \mathbb{I}[h_{0j}=y_j \land a_j\neq y_j \land h_{1j}\neq a_j]
    \]
    \item \textbf{Near-miss rate}: proportion of high-risk cases where an incorrect AI recommendation was narrowly avoided through human intervention or override.    
    \[
    \frac{1}{N}\sum_{j=1}^N 
    \mathbb{I}[
    a_j \neq y_j 
    \land h_{1j}=y_j 
    \land \mathrm{Risk}_j = \text{high}
    ]
    \]
\end{itemize}

\paragraph{Governance-in-use signals.}
These metrics operationalize governance as observable behavior in practice rather than documentation. Let $R_j$ indicate a rollback, $E_j$ an escalation, and $\pi_j$ a policy rule for case $j$.
\begin{itemize}
    \item \textbf{Rollback rate}: frequency of reversing AI-influenced decisions after review.
    \[
    \frac{1}{N}\sum_{j=1}^N \mathbb{I}[R_j=1]
    \]
    \item \textbf{Escalation rate}: proportion of cases referred for human or institutional oversight.
    \[
    \frac{1}{N}\sum_{j=1}^N \mathbb{I}[E_j=1]
    \]
    \item \textbf{Rule--behavior contradiction}: violations where required actions (e.g., escalation) are not taken.
    \[
    \frac{1}{N}\sum_{j=1}^N
    \mathbb{I}[\pi_j=\text{escalate} \land E_j=0]
    \]
\end{itemize}

% -----------------------
\subsection{Learning \& Readiness Metrics (What changed over time?)}

These metrics assess whether onboarding produces durable, transferable user capability.

\begin{itemize}
    \item \textbf{Calibration gap}: misalignment between reported confidence and actual correctness.
    \[
    \frac{1}{N}\sum_{j=1}^N |c_j-\mathbb{I}[h_{1j}=y_j]|
    \]
    \item \textbf{Reliance slope over time (behavioral calibration)}: change in sensitivity of agreement to AI correctness across sessions/blocks.
    For block/session $k$:
    \[
    \mathrm{Slope}_k=\Pr(h_1=a\mid a=y)_k-\Pr(h_1=a\mid a\neq y)_k
    \]
    and learning can be summarized as $\Delta \mathrm{Slope}=\mathrm{Slope}_{post}-\mathrm{Slope}_{pre}$ 
    \item \textbf{Retention}: stability of calibration-related metrics across multiple sessions or time intervals.
    \[
    |\mathrm{Calib}_{\text{session }k}-\mathrm{Calib}_{\text{session }k+1}|
    \]
    \item \textbf{Transfer}: consistency of performance or reliance across tasks, datasets, or model versions.
    \[
    |\mathrm{Metric}_{\text{task A}}-\mathrm{Metric}_{\text{task B}}|
    \]
\end{itemize}

\begin{table*}[t]
\centering
\small
\setlength{\tabcolsep}{6pt}
\renewcommand{\arraystretch}{1.35}
\caption{Mapping metrics to observable data sources, U--C--I onboarding stages, and the corresponding design actions they enable. (Part 1)}
\label{tab:metric_mapping_uci_part1}
\begin{tabular}{p{0.23\linewidth} p{0.24\linewidth} p{0.12\linewidth} p{0.33\linewidth}}
\toprule
\textbf{Metric (example)} & \textbf{Data source (trace)} & \textbf{U--C--I stage} & \textbf{Design action (what you do when it’s bad)} \\
\midrule

\multicolumn{4}{l}{\textbf{Outcome metrics (What happened?)}}\\
\midrule
\textbf{Team accuracy / TeamGain} & Final decision $h_1$, ground truth $y$; condition logs & Improve & Adjust delegation policy (when to defer / override); revise workflow to route high-risk cases to human review.\\
\textbf{Regret\_best / Oracle gap} & $h_0, a, h_1, y$ per case & Improve & Diagnose collaboration failures (not model limits); target training on cases where either human or AI was correct but the team failed.\\
\textbf{Error recovery vs.\ amplification} & $h_0 \rightarrow h_1$ transitions + $y$ & Improve & Identify whether UI causes harmful flips; refine prompts/explanations to reduce changed-to-wrong; add guardrails for high-stakes edits.\\

\midrule
\multicolumn{4}{l}{\textbf{Reliance \& interaction metrics (How was AI used?)}}\\
\midrule
{\parbox[t]{\linewidth}{\textbf{Accept-on-wrong}\\ \textbf{Reject-on-wrong}}}
& Agreement with AI ($h_1=a$) conditioned on AI correctness & \parbox[t]{\linewidth}{Understand\\[2pt]+ Control} & Curate ``failure sets''; add reliability cues; introduce regions-of-no-use; require justification or second-check for high-risk acceptance.\\
{\parbox[t]{\linewidth}{\textbf{ChangedToWrong}\\ \textbf{ChangedToRight}}} & Decision-change events ($h_1\neq h_0$) + $y$ & \parbox[t]{\linewidth}{Understand\\[2pt]+ Control} & Refine onboarding tasks to expose boundary conditions; add counterfactual practice; redesign explanation timing to prevent harmful flips.\\
\textbf{Override rate + timing} & Override events + timestamps $t_j$ & Control & Add safe levers (sandbox preview, rollback); reduce friction for appropriate overrides; introduce escalation shortcuts for uncertain cases.\\
\textbf{Reliance slope} & $\Pr(h_1=a\mid a=y)-\Pr(h_1=a\mid a\neq y)$ & Understand & Diagnose calibration; if low, strengthen training on discriminating correct vs.\ incorrect AI; improve uncertainty communication.\\
\textbf{Intervention latency} & $(t^{confirm/override}-t^{AI})$ & Control & Tune interaction costs; add ``pause-and-check'' for risky cases; streamline override/escalation to reduce delayed intervention.\\
%\midrule

\bottomrule
\end{tabular}
\end{table*}

\begin{table*}[t]
\centering
\small
\setlength{\tabcolsep}{5pt}
\renewcommand{\arraystretch}{1.35}
\caption{Mapping metrics to observable data sources, U--C--I onboarding stages, and the corresponding design actions they enable. (Part 2)}
\label{tab:metric_mapping_uci_part2}

\begin{tabular}{p{0.22\textwidth} p{0.23\textwidth} p{0.13\textwidth} p{0.34\textwidth}}
\toprule
\textbf{Metric (example)} & \textbf{Data source (trace)} & \textbf{U--C--I stage} & \textbf{Design action (recommended response)} \\
\midrule

\multicolumn{4}{l}{\textbf{Safety \& harm metrics (What went wrong?)}} \\
\midrule

\textbf{AI-harm / AI-help}
& Help--harm decomposition from $h_0, h_1, a, y$
& Control + Improve
& Add guardrails where AI induces harm; adjust autonomy (e.g., limit deferral) in high-harm regions; prioritize model fixes for harm-heavy slices. \\

\textbf{Missed-help / Under-reliance}
& $h_0 \neq y$, $a = y$, but $h_1 \neq a$
& Understand
& Improve ``when to trust'' instruction; show exemplars of correct AI behavior; add calibrated confidence cues for beneficial reliance. \\

{\parbox[t]{\linewidth}{\textbf{Near-misses}\\(high-risk disagreements)}}
& High-stakes disagreement logs; risk labels; margin/uncertainty if available
& Control
& Trigger required second review; add risk-based escalation rules; refine ``regions-of-no-use'' policy. \\

\textbf{Rule--behavior contradiction rate}
& Policy label $\pi_j$ + behavior event ($E_j$)
& Control + Improve
& Fix workflow compliance gaps; redesign the UI to make required actions salient; update governance policy or training based on observed violations. \\

\textbf{Rollback rate / Escalation rate}
& Rollback logs $R_j$; escalation logs $E_j$
& Control + Improve
& Audit contested decisions; improve contestability pathways; adjust accountability and when rollback is encouraged. \\

\midrule

\multicolumn{4}{l}{\textbf{Learning \& readiness metrics (What changed over time?)}} \\
\midrule

\textbf{Calibration gap}
& Confidence $c_j$ + correctness $\mathbb{I}[h_1 = y]$
& Understand
& Improve calibration cards; add feedback on confidence miscalibration; emphasize boundary conditions and failure modes. \\

\textbf{Retention}
& Same metrics across sessions (pre/post; follow-up)
& Improve
& Iterate the onboarding curriculum; schedule refreshers; adapt materials to failure modes that ``do not stick.'' \\

{\parbox[t]{\linewidth}{\textbf{Transfer}\\(across tasks/versions)}}
& Metrics across task A vs.\ B, or model version $v$ vs.\ $v'$
& Improve
& Update onboarding for new model versions; add regression tests for reliance and harm; retrain users on newly emerging failures. \\

{\parbox[t]{\linewidth}{\textbf{Time-to-calibration}}}
& Rolling-window estimates of reliance slope / accept-on-wrong over time
& \parbox[t]{\linewidth}{Understand\\[2pt]+ Improve}
& Personalize onboarding length; stop training when stable calibration is achieved; allocate extra practice for slow-to-calibrate users. \\

\bottomrule
\end{tabular}
\end{table*}


\end{document}